\documentclass[twocolumn,twoside,slac_two]{revtex4}
\usepackage{graphicx}
\usepackage{fancyhdr}
\usepackage{longtable}
\begin{document}

\title{Search for GeV gamma-ray emission from clusters of galaxies 
studied by TeV telescopes}

%

\author{Masaki Mori}
\affiliation{Department of Physics, College of Science and Engineering, 
Ritsumeikan University, Kusatsu, 525-8577 Shiga, Japan
(E-mail: {\tt morim@fc.ritsumei.ac.jp})}

\begin{abstract}
A cluster of galaxies is a huge system bounded by gravitation, 
and cosmic rays are thought to be confined in
the system, thus it should contain much non-thermal components. 
Many theories predict significant gamma-ray
emission that could be detectable by state-of-the-art gamma-ray 
telescopes. Some clusters have already
been studied by using {\sl Fermi} gamma-ray space telescope in the GeV 
band and Cherenkov telescopes in the
TeV band, but most clusters are not studied in both energy bands. 
Here I present results on GeV gamma-ray
emission from clusters of galaxies which have been given upper 
limits by Cherenkov telescopes using
{\sl Fermi} archival data.
\end{abstract}

\maketitle

\thispagestyle{fancy}


\section{INTRODUCTION}
A cluster of galaxies is a huge system bounded by gravitation, 
and is thought to be an ideal site for cosmic-ray acceleration,
since cosmic-rays are confined in the system for a long time
(see, e.g.\ Aharonian \cite{Aharonian2004}). 
High-energy particles, accelerated at large-scale shockes
associated with accretion and merger processes, or in other
processes, interact with ambient matter and
radiation fields to produce non-thermal emission from radio
to gamma-ray energies.

Gamma-ray emission in clusters may come from several 
processes (see, e.g.\ review by Blasi et al. \cite{Blasi2007}).
Collision of hign-energy protons with intra-cluster medium
produce gamma-ray emission via decay of neutral pions.
High-energy electrons can upscatter ambient photons
such as cosmic microwave background (CMB), infrared, starlight
and other soft photon fields to gamma-ray energies.
Gamma-ray production from inverse Compton scattering
by secondary electrons generated when ultra-high-energy
protons interact with CMB via $p\gamma \rightarrow 
p e^- e^+$ process. Dark matter annihilation could also
be a source of gamma-rays.

In the GeV band, 
association between clusters of galaxies and EGRET
untidentified sources have been claimed by
Totani \& Kitayama \cite{Totani2000}, Colafrancesco 
\cite{Colafrancesco}, Kawasaki \& Totani \cite{Kawasaki2002},
and Scharf \& Mukherjee \cite{Scharf2002},
but Reimer et al.\  \cite{Reimer2003} could not confirm
the association and gave upper limits for 58 clusters.
At TeV energies, there are several attempts to
search for emission from nearby clusters, but so far 
only upper limits have been reported for
the Perseus and Abell 2029 clusters with the Whipple
telescope \cite{Perkins2006}, Perseus with MAGIC 
\cite{Aleksic2009}, Abell 496 and Abell 85
with H.E.S.S.\ \cite{Aharonian2008}, Coma with
H.E.S.S.\ \cite{Aharonian2009}, Abell 3667 and 
Abell 4038 with CANGAROO-III \cite{Kiuchi2009}
(see Table \ref{tab:TeV}).

\begin{table*}[t]
\caption{Summary of TeV observations of clusters of galaxies. Upper
limits are given in unit of the Crab nebula flux in the same
energy range: $F(>E)\simeq 2\times 10^{-11}(E/1\,{\rm TeV})^{-1.6}$\,cm$^{-2}$s$^{-1}$.}
\begin{tabular}{|l|l|c|l|l|}
\hline \textbf{Name} & \textbf{Position} & \textbf{Redshift} &
\textbf{Upper Limit} & \textbf{Group \& Reference}
\\
\hline Perseus & 03h19m, $+41^\circ30'$ & 0.018 & $<13$\% Crab & Whipple \\
       &       &       & ($>400$~GeV, $0.3^\circ$) & Perkins et al. 2006
        \cite{Perkins2006} \\
\cline{4-5}
       &  &  & $<1\sim2$\% Crab & MAGIC \\
       &       &       & ($>150$~GeV, point-like) & Aleksic et al. 2009
       \cite{Aleksic2009} \\
\hline Abell 2029 & 15h10m, $+05^\circ45'$ & 0.077 & $<14$\% Crab & Whipple \\
       &       &       & ($>400$~GeV, $0.3^\circ$) & Perkins et al. 2006
       \cite{Perkins2006} \\
\hline Abell 496 & 04h34m, $-13^\circ16'$ & 0.033 & $<5$\% Crab & H.E.S.S. \\
       &       &       & ($>0.57$~TeV, $0.6^\circ$) & Aharonian et al. 2009
       \cite{Aharonian2008} \\
\hline Abell 85 & 00h42m, $-09^\circ21'$ & 0.055 & $<2$\% Crab & H.E.S.S. \\
       &       &       & ($>0.46$~TeV, $0.49^\circ$) & Aharonian et al. 2009
       \cite{Aharonian2008} \\
\hline Coma & 12h59m, $+27^\circ58'$ & 0.023 & $<15$\% Crab & H.E.S.S. \\
       &       &       & ($>1$~TeV, $0.4^\circ$) & Aharonian et al. 2009
       \cite{Aharonian2009} \\
\hline Abell 3667 & 20h12m, $-56^\circ50'$ & 0.055 & $<29$\% Crab & CANGAROO-III \\
       &       &       & ($>950$~GeV, $0.4^\circ$) & Kiuchi et al. 2009
       \cite{Kiuchi2009} \\
\hline Abell 4038 & 23h47m, $-28^\circ12'$ & 0.029 & $<12$\% Crab & CANGAROO-III \\
       &       &       & ($>750$~GeV, $0.25^\circ$) & Kiuchi et al. 2009
       \cite{Kiuchi2009} \\
\hline
\end{tabular}
\label{tab:TeV}
\end{table*}

Recently, Bechtol et al.\ \cite{Bechtol2009} reported results on 15
clusteres which are top-ranked by Pfrommer \cite{Pfrommer2008} using
the {\sl Fermi} Gamma-ray Space Telescope, but most of these cluster
samples do not overlap with clusters observed at TeV
energies. Here I report on a search for gamma-ray emission
in the GeV band with {\sl Fermi} for clusters which are already
observed by TeV telescopes as above, in order to increase
multiwavelength coverage of emission from clusters
for further discussion on their high-energy behavior.

\section{ANALYSIS}

{\sl Fermi} archival data were extracted from Fermi Science
Support Center and analyzed using provided tools
(Fermi Science Tools v.9.15.2).
Energy ranges used in the present analysis are from 200 MeV
to 10 GeV, and data periods are from 2008 August to 2009 September
for all the analyzed targets listed in Table \ref{tab:TeV}.
Only `diffuse' class events were used for analysis.

\begin{figure}[htbp]
\centering
\includegraphics[width=115mm]{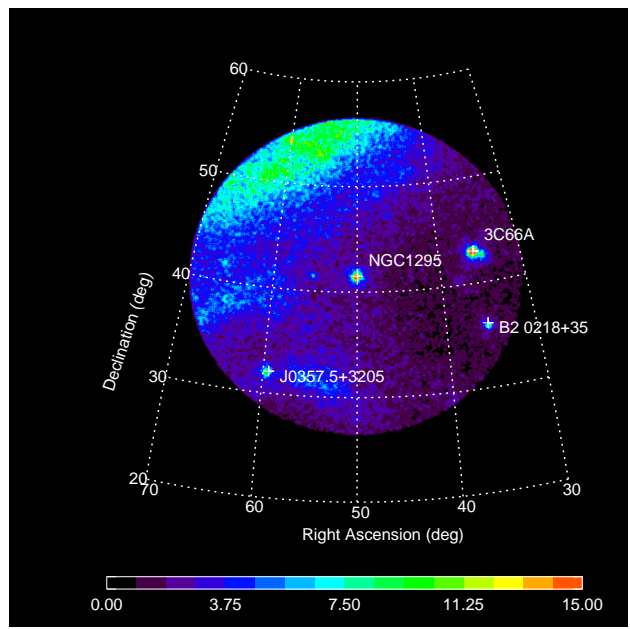}
\caption{Count map of the Perseus cluster, smoothed over neighboring
bins of $0.1^\circ$ square. (Smoothing is also applied for other count maps).} 
\label{fig:Perseus}
\end{figure}

\begin{figure}[htbp]
\centering
\includegraphics[width=115mm]{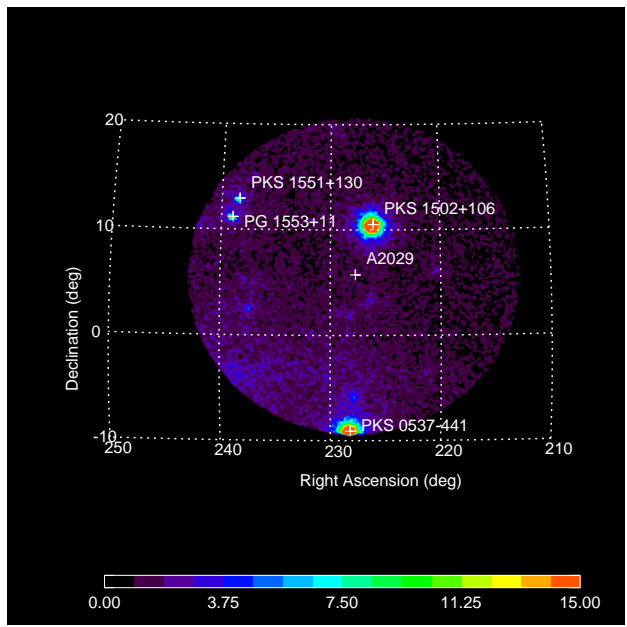}
\caption{Count map of Abell 2029.} 
\label{fig:A2029}
\end{figure}

\begin{figure}[htbp]
\centering
\includegraphics[width=115mm]{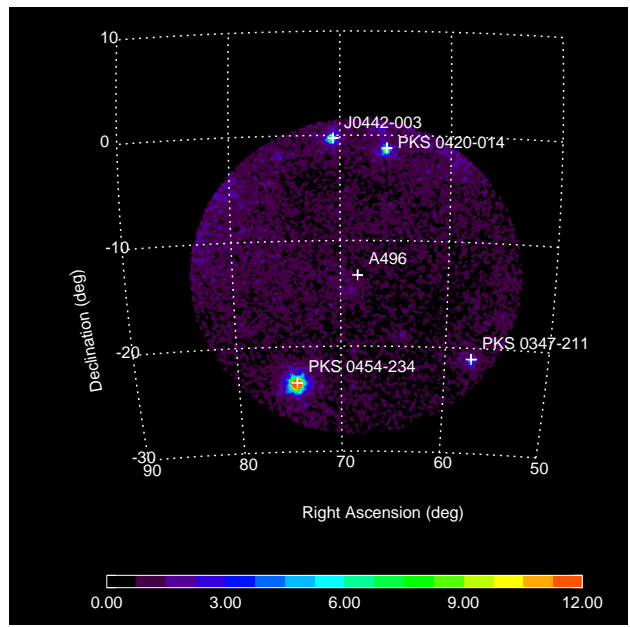}
\caption{Count map of Abell 496.} 
\label{fig:A0496}
\end{figure}

\begin{figure}[htbp]
\centering
\includegraphics[width=115mm]{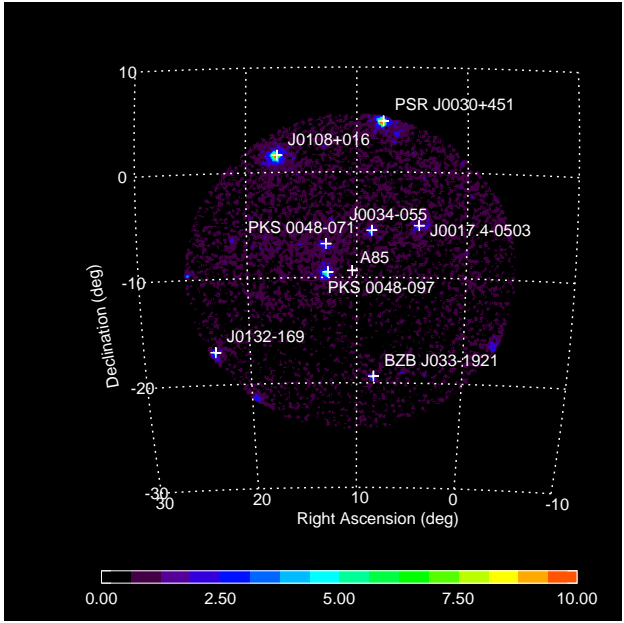}
\caption{Count map of Abell 85.} 
\label{fig:A0085}
\end{figure}

\begin{figure}[htbp]
\centering
\includegraphics[width=115mm]{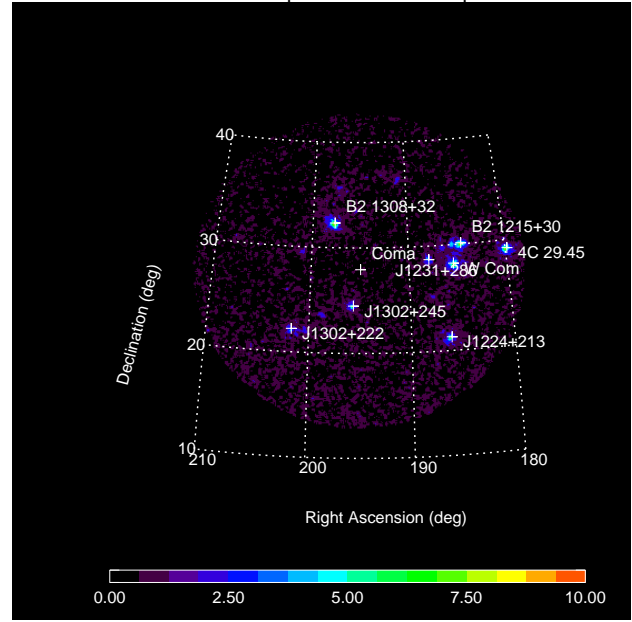}
\caption{Count map of the Coma cluster.} 
\label{fig:Coma}
\end{figure}

\begin{figure*}[t]
\centering
\begin{minipage}[t]{.48\textwidth}
\includegraphics[width=115mm]{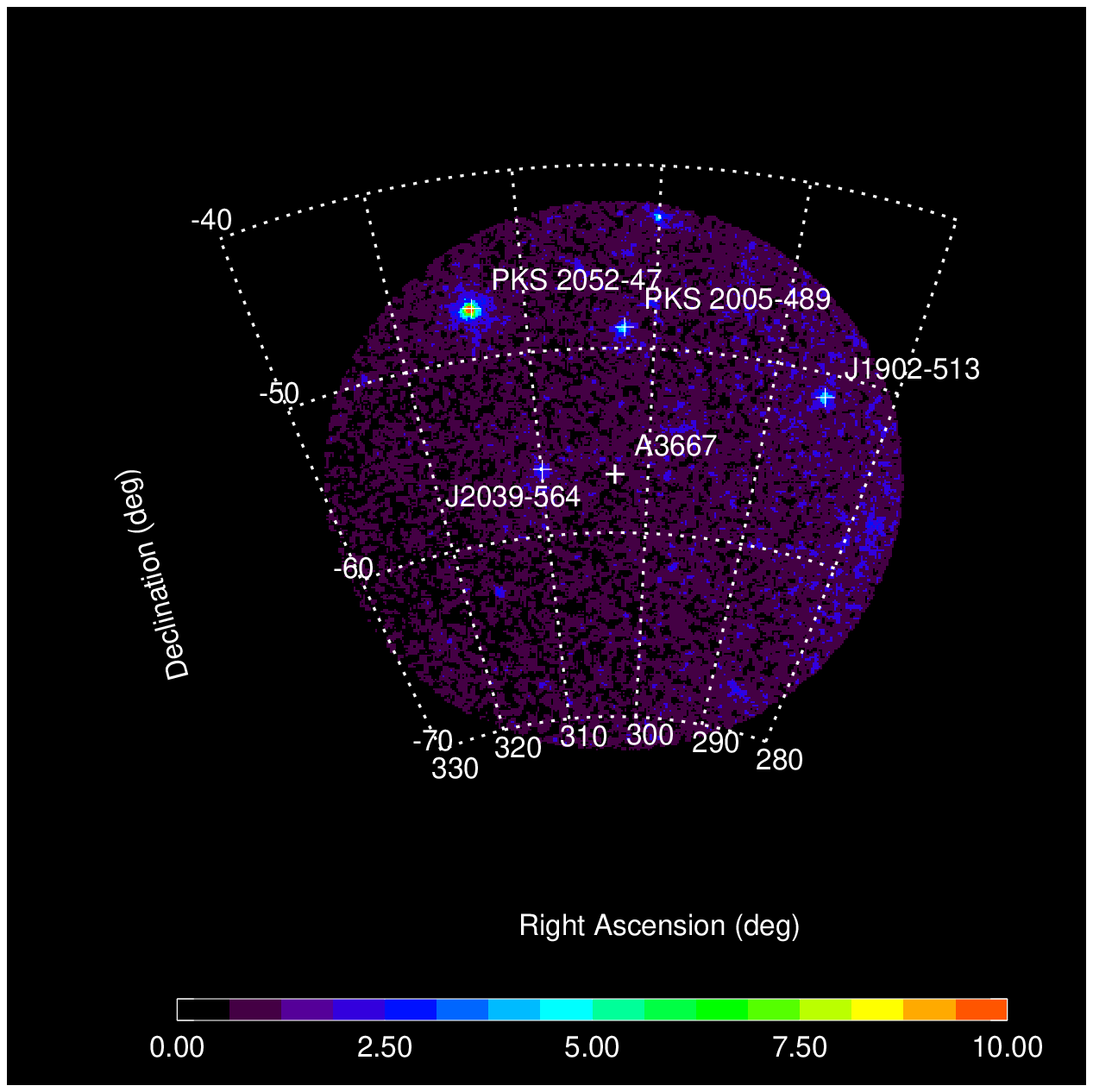}
\caption{Count map of Abell 3667.} 
\label{fig:A3667}
\end{minipage}
\hfill
\begin{minipage}[t]{.48\textwidth}
\includegraphics[width=115mm]{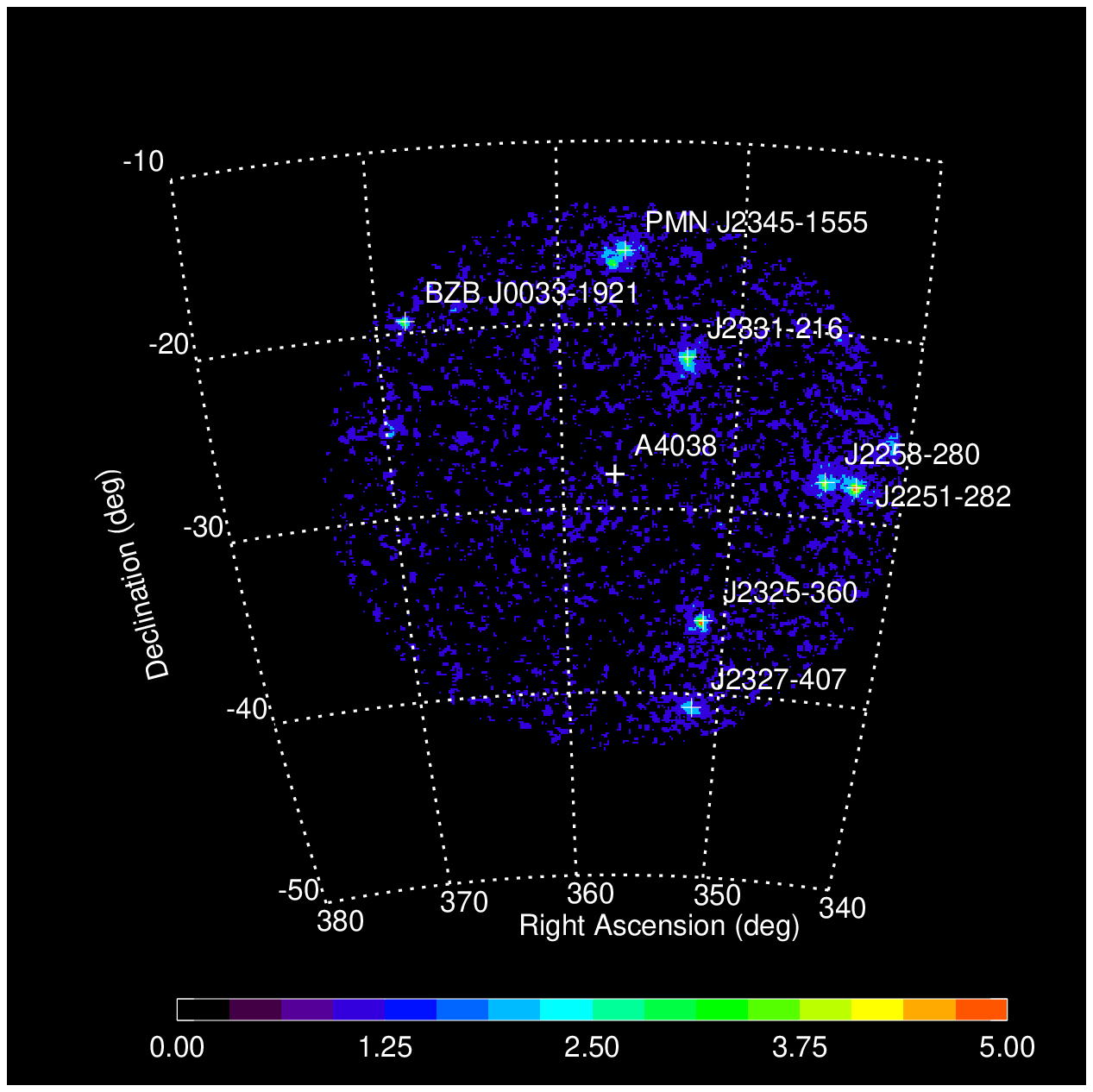}
\caption{Count map of Abell 4038.} 
\label{fig:A4038}
\end{minipage}
\end{figure*}

Figures \ref{fig:Perseus} to \ref{fig:A4038} are count maps
for each targets.
Except Perseus, no significant emission was found in any
of these objects. There is a point-like source in the
Perseus cluster, which is identified as NGC 1275 and
already reported in detail by Abdo et al.\ \cite{Abdo2009}.

%



The upper limits on gamma-ray emission from each cluster have
been calculated with a likelihood fitting program, {\it gtlike},
in unbinned mode, and assuming point-like sources at the
center of each clusters using `PowerLaw2' model so that
the limits are not sensitive to the power-law indices.
Sources listed in the {\sl Fermi} Bright Gamma-ray Source list 
\cite{Abdo2009a} and contained in the $15^\circ$-radius 
field of view are modeled in the {\it gtlike} analysis.
Prominent sources in the field, which
are not listed in Ref.\ \cite{Abdo2009a} but may affect the fitting
procedure due to the point spread function,
are identified by eye and also included in the analysis.
The profile likelihood method was used to calculate 95\% 
confidence level upper limits on gamma-ray flux by
setting $\chi^2-\chi_{\rm min}^2 = \Delta \chi^2=-2\Delta \ln{\cal L}=3.84$.
Results are summarized in Table \ref{tab:ul}.

\begin{table*}[tbp]
\caption{Summary of upper limits (95\% C.L.) on gamma-ray fluxes from 
clusters of galaxies. 
(Unit: $10^{-8}$cm$^{-2}$s$^{-1}$)}
\begin{tabular}{|l|c|c|c|l|}
\hline \textbf{Name} & \textbf{Upper Limit}  & \textbf{Upper Limit} &
\textbf{Bechtol et al.} & \textbf{Note}
\\
                     & ($>100$~MeV) & ($>200$~MeV) & ($>100$~MeV) & 
\\
\hline Perseus    & --   & --     & 20 & NGC 1275 (point source) \\ 
\hline Abell 2029 & 4.8  & 1.3    & -- &  \\
\hline Abell 496  & 1.2  & 0.61   & -- &  \\
\hline Abell 85   & 0.12 & 0.062  & -- &  \\
\hline Coma       & 0.88 & 0.44   & 0.6 &  \\
\hline Abell 3667 & 0.23 & 0.095  & -- &  \\
\hline Abell 4038 & 0.52 & 0.45   & -- &  \\
\hline
\end{tabular}
\label{tab:ul}
\end{table*}

\section{DISCUSSION}

\begin{figure}[htbp]
\centering
\includegraphics[width=90mm]{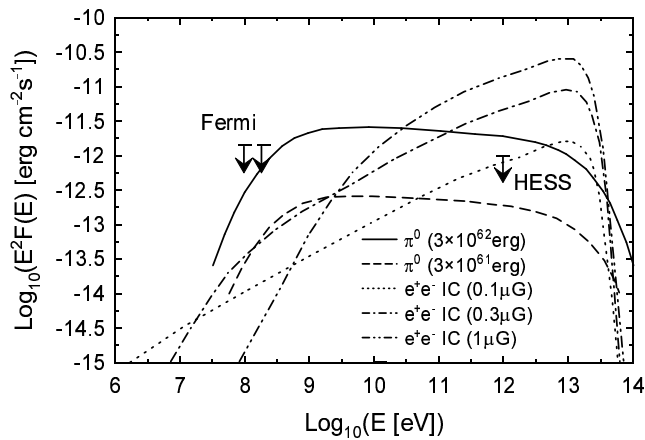}
\caption{Model predictions and upper limits on the gamma-ary emission
from the Coma cluster.  $\pi^0$
model curves are taken from V\"{o}lk and Atoyan \cite{Voelk2000}
and e$^+$e$^-$ IC from Inoue, Sugiyama and 
Aharonian \cite{Inoue2005}.} 
\label{fig:sed}
\end{figure}

Since no significant emission was found from any of 7 clusters of galaxies
(other than Perseus), we discuss restrictions
on some emission models. Here we take the Coma cluster as an example.
Figure \ref{fig:sed} is a comparison of experimental limits
on gamma-ray fluxes with some model predictions for the
Coma cluster case. $\pi^0$
model curves are taken from V\"{o}lk and Atoyan \cite{Voelk2000}
and e$^+$e$^-$ IC model curves are from Inoue, Sugiyama and 
Aharonian \cite{Inoue2005}.  Upper limits based on
GeV observations are just about to constrain $\pi^0$ model
predictions, but do not restrict e$^+$e$^-$ IC models within a
plausible range of parameters.
Although the TeV upper limit is a little more restrictive
on the  $\pi^0$ model, accumulation of {\sl Fermi} data over a few years
may reveal hadronic emission from the Coma cluster if the maximum
energy of accelerated protons are lower than the assumed
model or the spectrum is steeper.

\bigskip 
\begin{acknowledgments}
This work is supported by Ritsumeikan University Research Fund.
\end{acknowledgments}

\bigskip 

\end{document}